\documentclass[12pt,tighten,flushrt,preprint]{aastex}

 \usepackage{apjfonts,emulateapj5}

\newcommand{\kms}{km~s$^{-1}$}
\newcommand{\cps}{ct~s$^{-1}$}
\newcommand{\ergps}{erg~s$^{-1}$}
\newcommand{\flx}{ph~cm$^{-2}$~s$^{-1}$}
\newcommand{\hr}{\mbox{$^{\mathrm h}$}}
\newcommand{\mn}{\mbox{$^{\mathrm m}$}}

\slugcomment{To appear in the Astrophysical Journal}

\shorttitle{X-Ray Emissions of UV Ceti A and B}
\shortauthors{Audard, G\"udel, \& Skinner}

\begin{document}

\title{Separating the X-Ray Emissions of UV Ceti A and B with \textit{Chandra}}

%
%
%

\author{Marc Audard\altaffilmark{1}, Manuel G\"udel\altaffilmark{2}, and Stephen L. Skinner\altaffilmark{3}}

\altaffiltext{1}{Columbia Astrophysics Laboratory, Mail code 5247, 550 West
120th Street, New York, NY 10027, {audard@astro.columbia.edu}}
\altaffiltext{2}{Paul Scherrer Institut, Villigen \& W\"urenlingen, 5232 Villigen PSI, Switzerland, {guedel@astro.phys.ethz.ch}}
\altaffiltext{3}{Center for Astrophysics and Space Astronomy, University of Colorado, Boulder, CO 80309-0389, {skinners@casa.colorado.edu}}

\begin{abstract}
We present the first spatially separated X-ray observation of the active UV Ceti 
binary with \textit{Chandra}, providing important new information on 
the relative X-ray activity levels of its nearly identical dM5.5e components. 
The two components had similar X-ray luminosities during low-level periods
and emission measure loci derived from Low Energy Transmission
Grating spectra reveal similar coronal temperatures ($3 - 6$~MK). However, 
the B component showed a higher degree of variability, resulting
in a higher average X-ray luminosity than that of UV Cet A. 
We discuss possible causes of this enhanced activity in UV Ceti B.
\end{abstract}


\section{Introduction}

The nearby ($d = 2.7$~pc) binary \objectname{UV Ceti}
(\mbox{\objectname{GJ~65~AB}}; 
\mbox{\objectname{L726~-8~AB}}) is composed of two dM5.5e 
stars with similar masses and radii ($M \approx 0.1~M_\sun$, $R \approx 
0.15~R_\sun$) in a highly eccentric orbit with a period of 26.5~y and a
semi-major axis of $\approx 2\arcsec$ \citep*[$5.06$~AU;][]{worley73,geyer88}. 
UV Ceti is the prototype of flare stars and has been extensively studied
from radio to X-rays \citep*[e.g.,][]{dejager89,stepanov95,guedel96}. 
Previous X-ray observatories could not separate the components 
\citep*{pallavicini90,schmitt93}. However, high angular resolution radio 
observations have resolved the binary  into its components and show that 
their radio emissions significantly differ: UV Ceti B flares more frequently than UV 
Ceti A and is also an order of magnitude brighter in its steady radio emission
\citep[e.g.,][]{gary81,gary82,kundu87,jackson87,kundu88,jackson89,benz98}.
The different radio behavior between UV Cet A/B raises the
   obvious question of whether the X-ray properties of the two
   components are also different. Although previous X-ray
   observatories lacked sufficient angular resolution to answer
   this question, it can now be addressed with \textit{Chandra}. We present
   here the first spatially separated X-ray observation of
   UV Ceti A/B. Light curves and spectral emission lines were extracted 
   for both components. A detailed analysis of the 
\textit{Chandra} data, and of \textit{XMM-Newton} and VLA observations 
of UV Ceti will be presented in a forthcoming paper.

\section{Observations}
The \textit{Chandra} X-ray Observatory observed the UV Ceti binary for 75~ksec
from 2001 November 26 at 20:55:39 UT to 2001 November 27 at 18:09:27 UT, with the 
Low Energy Transmission Grating Spectrometer (LETGS) inserted. This configuration 
provides grating spectra from 1.7 to 177~\AA\  with a constant resolution of 
$\Delta\lambda \approx 0.05$~\AA\  (FWHM). The positive and negative 
order spectra are dispersed on the High-Resolution Camera (HRC-S) on each 
side of the focal aim point, and the latter provides a direct zeroth order image 
of the target in the energy range $0.08-10$~keV with an angular resolution of 
0.4\arcsec\  (FWHM). The  HRC ``pixel'' size is $0.13175\arcsec$. 
Further details on the instruments can be found in the 
\textit{Chandra} \anchor{http://cxc.harvard.edu/proposer/POG}{Proposers'
Observatory Guide}\footnote{\url{http://cxc.harvard.edu/proposer/POG}}.

We reduced level 2 files provided by the \textit{Chandra} X-ray Center,
following recommended procedures with the CIAO 2.2.1 software.
The binary was aligned almost along the dispersion axis, and was separated
in the grating spectra by about 80~m\AA.
This way, in the negative order spectrum, emission lines of UV Cet A were 
shifted to shorter wavelengths relative to UV Cet B, whereas they were shifted 
to longer wavelengths in the positive order. 

\section{Separating UV Ceti A and B}

\begin{figure}
\plotone{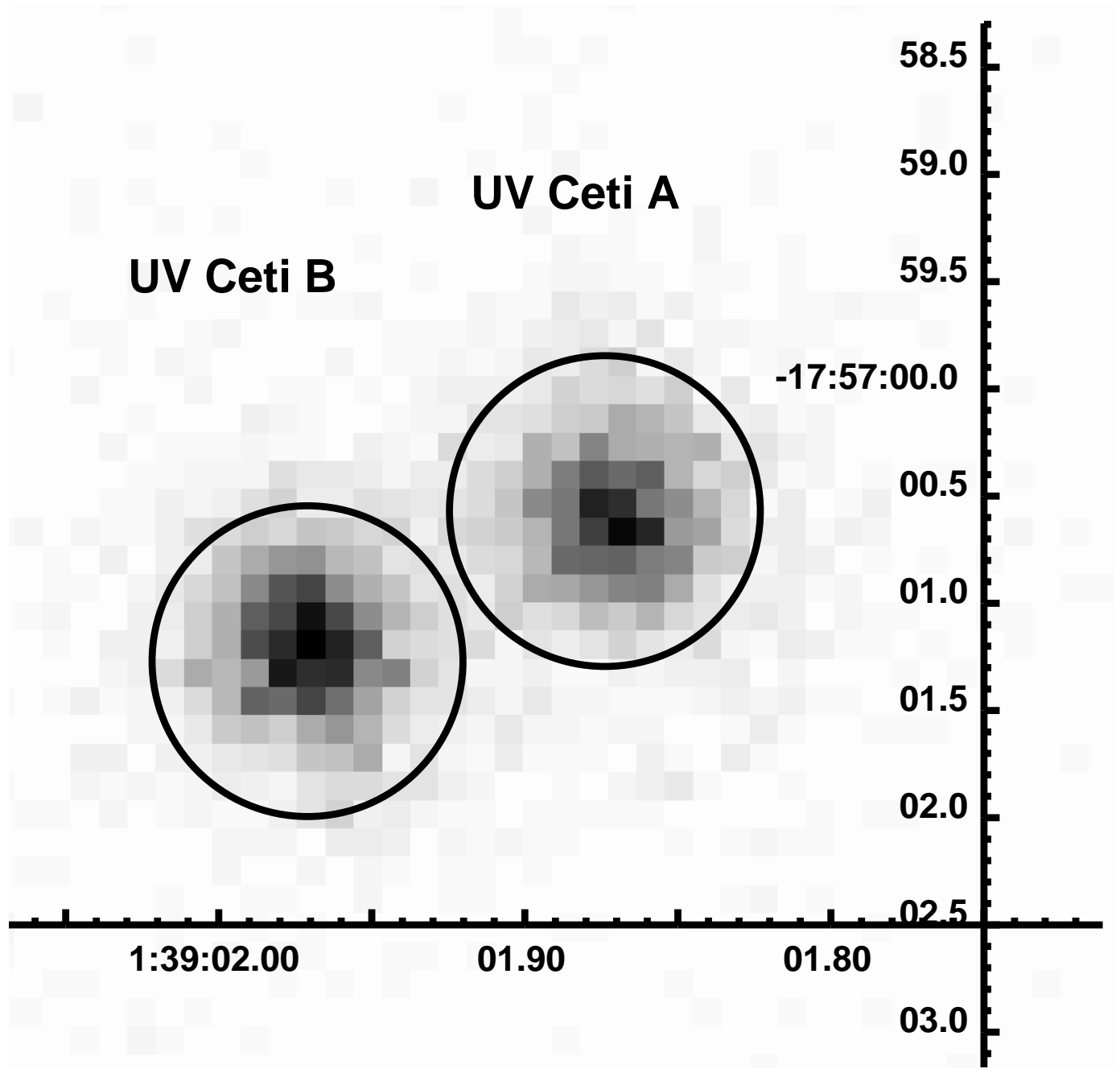}
\caption{Zeroth order image of the low-level X-ray emissions of UV Ceti A and B.
Circles of 5.5 pixels ($0.72\arcsec$) are centered on the positions 
of each component. The axes give the J2000 coordinates in right ascension (horizontal) 
and declination (vertical). The coordinates for UV Ceti A are $\alpha = 1\hr 39\mn 01\fs 87$ 
and $\delta = -17\arcdeg 57\arcmin 00\farcs 6$, whereas they are 
$\alpha = 1\hr 39\mn 01\fs 97$ and $\delta = -17\arcdeg 57\arcmin 01\farcs 3$ for UV Ceti B.
\label{fig1}}
\end{figure}

Figure~\ref{fig1} shows the zeroth order image of the binary; the 
astrometry is expected to be accurate within 1\arcsec.
We used a sliding cell source detection algorithm with a fixed 
cell size of 6 pixels ($0.79\arcsec$).
The components were separated by $1.54\arcsec$ with UV Ceti B lying at a 
position angle of $117.05\arcdeg$ relative to A. 
At the epoch of our observation, the orbital solution of \citet{geyer88} gives 
a separation of $d=1.45\arcsec$ and a position angle
of $117.1\arcdeg$, whereas the solution of \citet{worley73} gives 
$d=1.54\arcsec$ and an angle of $117.3\arcdeg$. Our X-ray positions are therefore in
excellent agreement with the expected optical positions.

\section{Light Curves}
We extracted zeroth order light curves for each component using circular
regions with radii of 5.5 pixels ($\approx 0.72\arcsec$) and applied dead-time
corrections. The background contribution was estimated from nearby regions 
and was negligible  ($\approx 5.6 \times 10^{-5}$~\cps, scaled to the source extraction
region).

Figure~\ref{fig2} shows the X-ray light curves for both components. Obviously,
UV Cet B was highly variable and showed at least five large flares, while
UV Cet A only showed a gradual rise and fall at the start of the
observation. The small flare in UV Cet A occurring simultaneously with the large flare in 
UV Cet B is an artifact due to flux contamination from component B to component A.
The two stars had a similar low-level X-ray count rate (excluding the five large flares,
 see Fig.~\ref{fig2}), around $0.025-0.03$~\cps\  ($L_\mathrm{X} 
\approx 1.0 - 1.2 \times 10^{27}$~\ergps\  for each component, 0.1-10~keV; based 
on a simulation of a 0.34~keV [see \S\ref{spectra}] coronal plasma with solar 
abundances in the Portable, Interactive Multi-Mission Simulator, PIMMS; \citealt{mukai93}).

UV Cet B displayed two types of flares: the first type is characterized by a 
slow increase and a slow decay, with similar time scales. The second flare type
is reminiscent of ``impulsive'' solar flares, with a rapid rise and a slower decay. 
Both flare types however share similar total durations of $\approx 2000-3000$~s. 
The peak count rate of the giant flare in UV Cet B is about a factor of 100
over the low-level count rate ($L_\mathrm{X,peak} \approx 2 \times 10^{29}$~\ergps; for a 
1.37~keV plasma in PIMMS); the flare has a very short rise time of 80~s, and
returns to the preflare level after $\approx 2300$~s (see inset in 
Fig.~\ref{fig2}). 
A secondary peak occurs during the decay phase of the first flare, perhaps indicating 
the occurrence of a second flare. 

\section{Spectra}\label{spectra}

We present in this section a discussion of the \textit{Chandra} LETGS spectra 
of UV Ceti A and B. In this paper, our analysis focuses on the low-level
emission (50.8~ksec), however we do note that emission lines from higher ionization stages
of Fe (e.g., \ion{Fe}{23} and \ion{Fe}{24} in the 10-12~\AA\  range) were seen 
during flares.
The selected time intervals for the low-level emission are 
shown in the upper panel of Fig.~\ref{fig2}.
Figure~\ref{fig3} shows the full \textit{Chandra} positive and negative order 
spectra during the low-level emission. Since the latter is faint (a total of 
0.11~\cps\  in the dispersed spectra), only a few bright emission 
lines have been positively detected. 
The low-level X-ray spectra of UV Cet A and B are
similar. They show the strongest emission in the \ion{O}{8}~Ly$\alpha$ line.
Emission lines from \ion{Fe}{17}, \ion{Ne}{9}, \ion{Ne}{10}, \ion{C}{6},
\ion{N}{7}, \ion{O}{7} are also detected below 40~\AA. The range of temperatures over 
which these lines form indicates that there is significant emission measure (EM) in the 
range $T = 3 - 6$~MK, which is confirmed by electron temperature diagnostics such as 
line ratios. 
Reliable density measurements could not be obtained because of low
fluxes in the closely-spaced He-like triplets and line blending (see, e.g., panels b 
and d in Fig.~\ref{fig4}).
At longer wavelengths ($\geq 40$~\AA), the signal-to-noise ratio was 
insufficient to significantly detect emission lines. We derived upper 
limits to fluxes of Fe L-shell lines with typical large emissivities, such
as \ion{Fe}{18} $\lambda$  93.92~\AA, \ion{Fe}{19} $\lambda$ 108.37~\AA, 
\ion{Fe}{21} $\lambda$ 128.73~\AA, \ion{Fe}{20}/\ion{Fe}{23} $\lambda$ 132.85~\AA, 
\ion{Fe}{9} $\lambda$ 171.07~\AA, and \ion{Fe}{10} $\lambda$ 174.53~\AA.
The absence of detectable emission from these lines suggests that there is no 
significant EM at high ($T \geq 6$~MK) or low ($T \leq 2$~MK) temperatures. 

We fitted the positive and negative order spectra simultaneously with two
monochromatic emission lines per feature (for UV Cet A and B) folded through the 
\textit{Chandra} line response function. In the negative order spectrum, the
model line wavelengths were shifted by $-12$~m\AA\  to account for the wavelength
discrepancy between the positive and negative order spectra with CIAO 2.2.1.
The emission lines from UV Cet A and B were separated by a fixed 80~m\AA.
First order grating
\anchor{http://cxc.harvard.edu/cal/Links/Letg/User/Hrc_QE/ea_index.html}{response
matrices from July 2002} as provided by 
the LETGS calibration team were used\footnote{\url{http://cxc.harvard.edu/cal/Links/Letg/User/Hrc\_QE/ea\_index.html}}.
Line emissivities from APEC 1.2 \citep{smith01} were used to calculate the EM
loci curves. 
We assumed that the coronal abundances in UV Cet A and B were identical to the
solar photospheric abundances \citep{anders89}. The line fluxes were not corrected 
for interstellar absorption, since it is very small ($N_\mathrm{H} 
\approx 6 \times 10^{17}$~cm$^{-2}$, based on the canonical interstellar density
$n_\mathrm{H} \approx 0.07$~cm$^{-3}$, \citealt{paresce84}),
even at the long wavelength end of the LETGS ($8$\% absorption at 170~\AA).
We provide EM loci of UV Cet A and B in Figure~\ref{fig5};
EM loci of detected Fe lines (solid curves) and of 
detected other elements (dotted curves) are shown with 68\% confidence ranges 
at the maximum formation temperatures ($T_\mathrm{max}$), suggesting that Ne and
O are overabundant relative to Fe. Upper limits (90\%
confidence ranges) for the EM loci of undetected Fe lines (dashed curves) 
are also shown with arrows at $T_\mathrm{max}$.

We have checked the consistency of our upper limits with the archival
\textit{EUVE} data of UV Cet. Since this paper aims at presenting the new
\textit{Chandra} results, we present here only a summary of the
\textit{EUVE} analysis. The UV Cet binary was observed during a time span of
885~ksec, with an effective exposure of about 230~ksec. Its Deep Survey 
(DS; 65--190~\AA) light curve displayed strong variability with several short
flare-like events. We extracted the Short Wavelength (SW; 70--190~\AA) spectrum
during the low-level emission (DS count rates lower than 0.1~\cps; exposure of
188~ksec), and compared the line fluxes with those in the \textit{Chandra} LETGS
spectrum. We emphasize that \textit{EUVE} could not separate 
the binary spatially and spectroscopically.
Only two lines were detected,
the \ion{Fe}{20}/\ion{Fe}{23} $\lambda$ 132.85~\AA\  blend, and the \ion{Fe}{18}
$\lambda$ 93.92~\AA\  line; their measured fluxes were $(8.1 \pm 5.0) \times
10^{-5}$~\flx\  and $(5.9 \pm 3.2) \times 10^{-5}$~\flx, respectively. 
The sum of the upper limits for UV Cet A/B obtained from the LETGS ($\lambda$ 
132.85~\AA: $\le 6.2 \times 10^{-5}$ and $\le 9.1 \times 10^{-5}$~\flx, 
$\lambda$ 93.92~\AA: $\le 5.4 \times 10^{-5}$ and $\le 6.2 \times 10^{-5}$~\flx, 
for UV Cet A and B, respectively) are compatible with the measured fluxes in
\textit{EUVE} SW.
The other Fe L-shell lines formed at high or low temperature (see 
previous paragraph) remain undetected by \textit{EUVE}, and their upper limits 
are also consistent with our \textit{Chandra} upper limits. 

\section{Discussion and Conclusions}
Numerous radio observations of the UV Ceti binary system have shown that
the coronae of both components differ significantly. The non-flaring radio emission
of the primary is much weaker than that of the secondary and often remains 
undetected \citep[e.g.,][]{gary82,kundu87}. Their flares also
are different: on UV Cet A, flares are often highly circularly polarized,
thus suggesting some coherent emission mechanism \citep{gary82,benz98,bingham01},
whereas flares in UV Cet B often show only weak or moderate polarizations
consistent with gyrosynchrotron emission \citep[e.g.,][]{jackson87,guedel96},
and occur more frequently than flares in the primary. 

\begin{figure}
\plotone{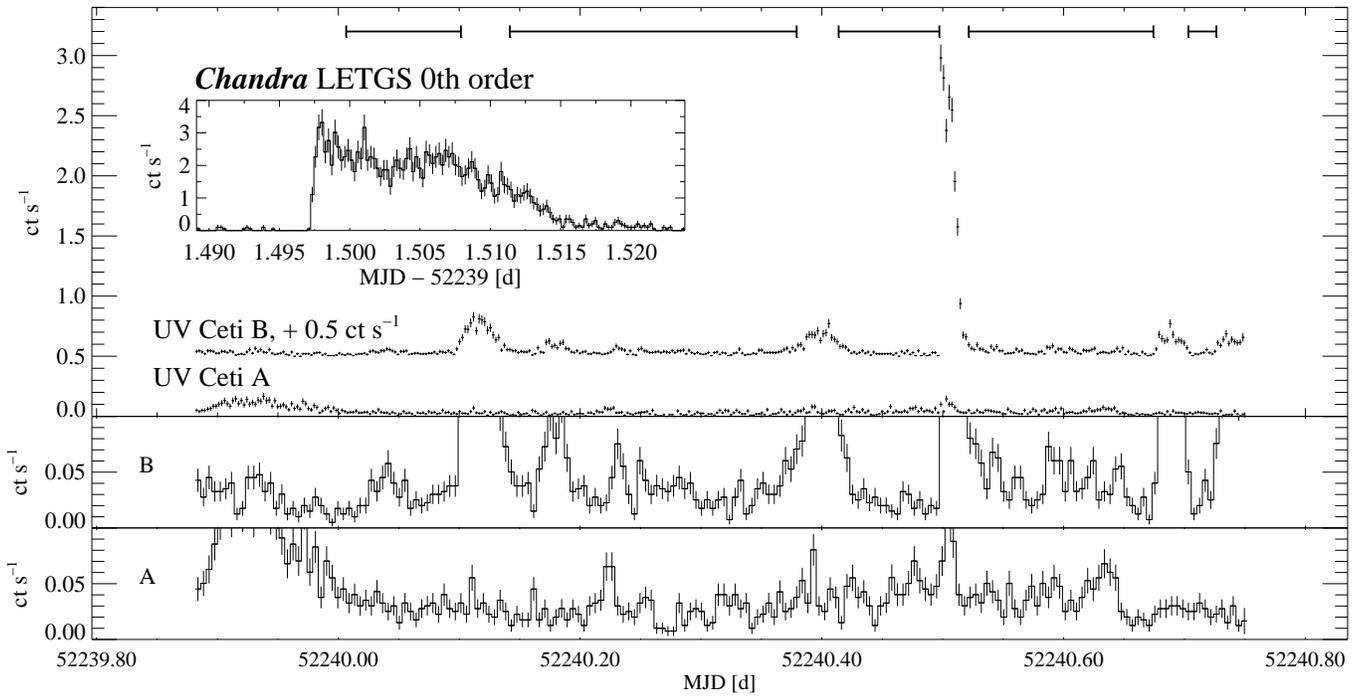}
\caption{Zeroth order HRC-S/LETG light curves of UV Ceti A and B (0.1-10~keV). 
\textit{Top panel:} The bin size is 200~s; the curve for UV Ceti B has been 
shifted by +0.5~\cps,  for clarity. The five time spans at the top represent the
selected time intervals for the low-level LETGS spectra shown in
Fig.~\ref{fig3}. \textit{Inset:} Extract of the large flare 
on UV Ceti B for a bin size of 20~s. The flare emission increased by a factor of 
about 100 above the low-level emission. The flare rise time is $
\approx 80$~s, whereas the count rate returns to the pre-flare level after 
$\approx 2300$~s. \textit{Middle and bottom panels:} Low-level 
X-ray emission of UV Ceti B (middle) and UV Ceti A (bottom), respectively. 
The bin size is 400~s.\label{fig2}}
\end{figure}

\begin{figure}
\plotone{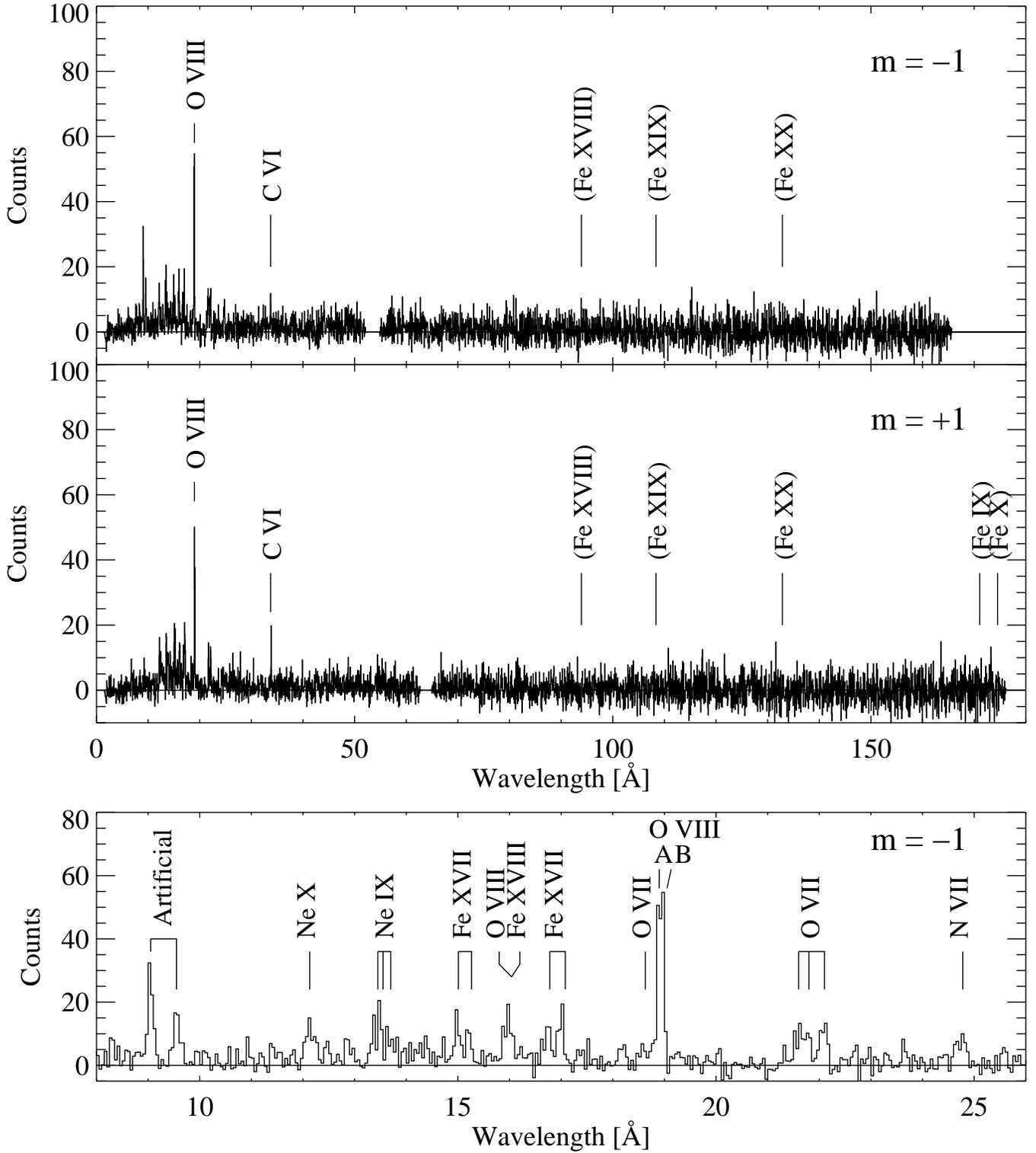}
\caption{\textit{Top and middle:} Low-level \textit{Chandra} LETGS spectra in
the negative (top) and positive (middle) orders for an exposure time of 50.8~ksec
and a binsize of 50~m\AA. With such a binsize, the contributions from each source
are averaged. We indicated in parentheses the expected
location of bright lines formed at high temperature (\ion{Fe}{18} -- \ion{Fe}{20}) 
and at low temperature (\ion{Fe}{9} and \ion{Fe}{10}) in the long wavelength range.
\textit{Bottom}: Extract of the negative order spectrum between 8 and 26~\AA. 
Two emission features at 9.04 and 9.55~\AA\  appear in the negative order
spectrum and have been identified as zeroth order contaminations from faint 
sources that fell, by chance, on the dispersed spectrum.\label{fig3}}
\end{figure}

\begin{figure}
\plotone{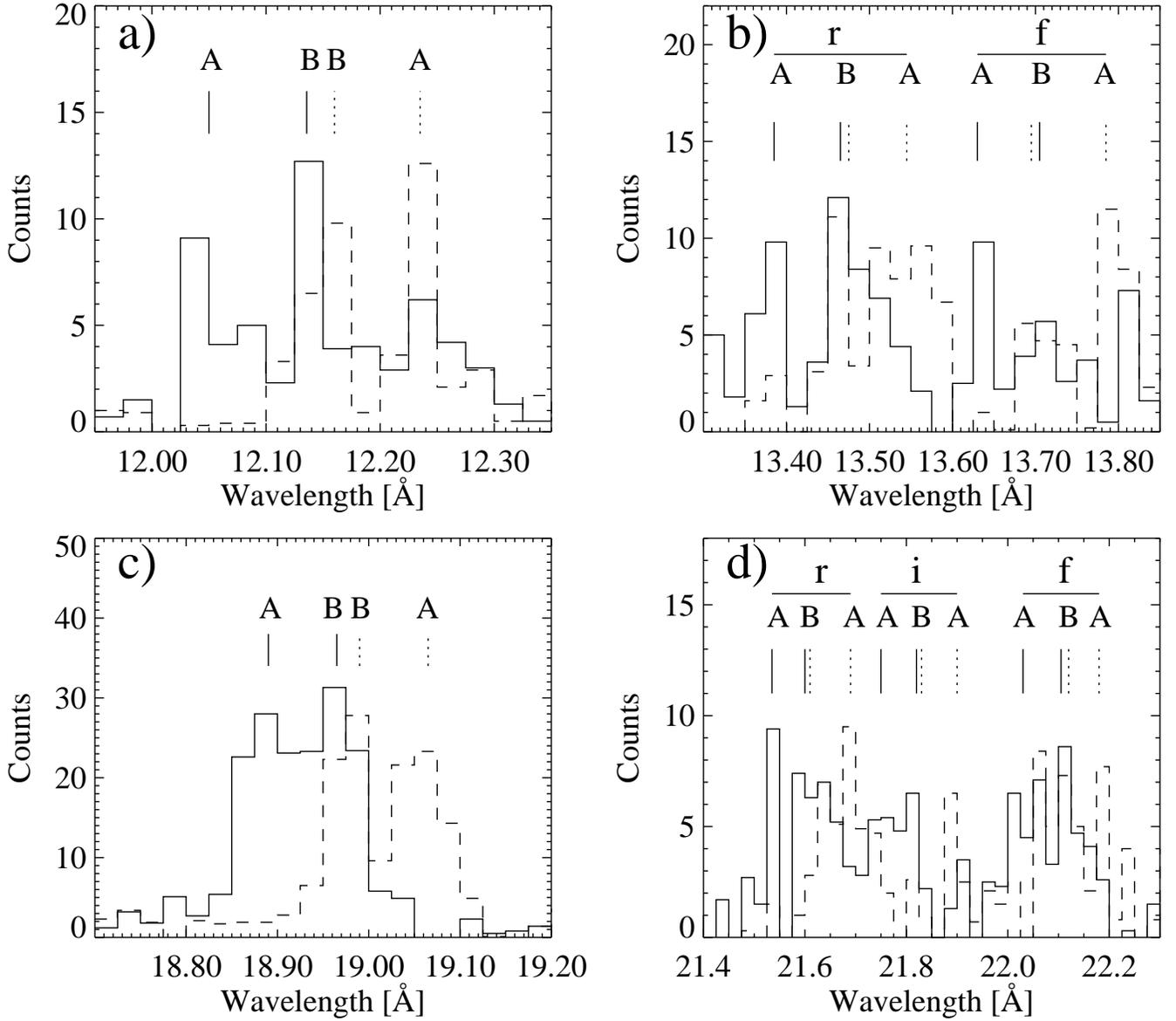}
\caption{Extract from low-level \textit{Chandra} spectra in the region of
\ion{Ne}{10}~Ly$\alpha$ (a), \ion{Ne}{9}~He$\alpha$ (b),
\ion{O}{8}~Ly$\alpha$ (c), and \ion{O}{7}~He$\alpha$ (d) for a binsize of
25~m\AA\  and an exposure time of 50.8~ksec. The solid line shows the negative 
order spectrum, whereas the dashed line stands for the positive order spectrum. 
The wavelength discrepancy originates from the 
inaccuracy of the relative wavelength calibration between the negative and 
positive order LETGS spectra with CIAO 2.2.1. In each panel, the emission of each binary component 
has been labeled. In panel (b), we show only the resonance ($r$) and forbidden 
($f$) lines, whereas in panel (d), the intercombination ($i$) line is also
shown.\label{fig4}}
\end{figure}

\begin{figure}
\plotone{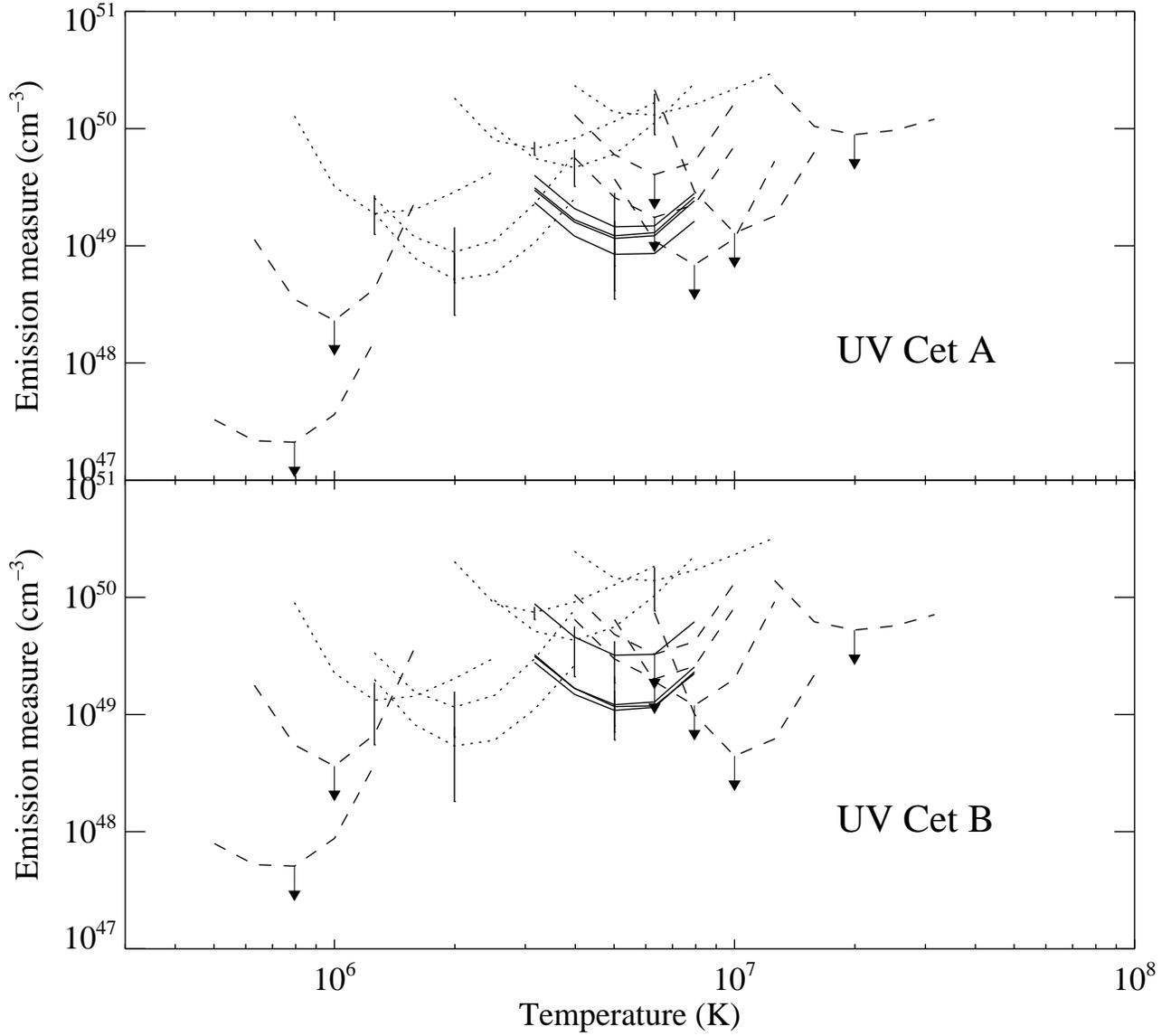}
\caption{Emission measure loci for the emission lines in UV Cet A and B.
Emissivities from APEC 1.2 were used assuming solar photospheric abundances
\citep{anders89}. The solid curves are for detected Fe lines (with 68\%
confidence ranges shown as vertical ranges), the dotted curves are for detected lines from other ions
(\ion{O}{8}, \ion{O}{7}, \ion{Ne}{10}, \ion{Ne}{9}, \ion{C}{6}, with 68\%
confidence ranges shown as vertical ranges), and the dashed curves are emission measure loci for the
upper limits (from 90\% confidence ranges) of undetected Fe lines.\label{fig5}}
\end{figure}

The contrasting behavior in the radio is peculiar since the two stars 
are otherwise thought to be very similar, with almost identical masses and 
radii. 
A study of their spatially separated X-ray emissions thus promises to uncover 
differences in the coronal energy release behavior of the two stars.
UV Cet A did not display any strong X-ray variability,
while UV Cet B was on average twice as luminous as the primary, essentially
because it displayed a higher flare rate (Fig.~\ref{fig2}, top panel). 
 The radio emission is due to the production  of high-energy electrons 
 during the flare process. Gyrosynchrotron emission would, under solar-like 
 conditions, decay rapidly due to radiation and collisional losses (within 
 minutes). However, if extended, weak magnetic fields were present, then part 
 of the relativistic electron population produced during the detected large
 X-ray flares could be efficiently trapped for durations much longer than 
 the characteristic time between large flares since
 $\tau \propto B^{-3/2}$ for synchrotron losses. Spatially
 resolved VLBA observations support such a model for UV Cet B (whereas
 no extended structure was identified in the primary). \citet{benz98}
 inferred magnetospheric structures around UV Cet B several times larger
 than the star, and from the decay time of a gradual flare ($\tau \approx 6500$~s)
 they inferred magnetic field strengths of only a few tens of Gauss. Based on
 this observation, \citet{kellett02} further proposed a magnetic trap model 
 that would at the same time explain steady radio and X-ray emission.

In contrast to the quasi-steady radio emission of each component,
the quasi-steady X-ray count rates 
of the two components are similar. A closer look at the \textit{Chandra} 
low-level light curve of UV Cet B (Fig.~\ref{fig2}, middle panel) shows that 
its ``quiescent'' emission is in fact flickering and is  composed of numerous 
temporally resolved flares, with a possible quasi-steady emission not exceeding 
0.01~\cps\  or 30\% of the average low-level emission. It may well be that the 
entire X-ray emission is a superposition of the detected flare decays, 
including very small events. This could be similar for the  X-ray emission of 
UV Cet A
(Fig.~\ref{fig2}, bottom panel). 
  Accelerated electrons from these flares can further add to the
  quasi-steady radio emission, but they are particularly well 
  trapped in the extended magnetosphere of UV Cet B.
\textit{XMM-Newton} data of the dM5.5e star Proxima Centauri have yielded similar 
results; its quiescent X-ray emission was temporally resolved into
a continual sequence of weak flares \citep{guedel02}. These results provide 
strong support for the hypothesis that low-energy flares contribute 
significantly to the coronal heating in stars \citep{audard00,kashyap02,guedel03}. 
In the Sun, flares with small energies contribute a significant amount to 
the total coronal heating \citep[e.g.,][]{krucker98,parnell00}.

In conclusion, 
the radio and X-ray behaviors of UV Cet A and
B are strikingly different. VLBA measurements have revealed an
extended magnetosphere around UV Cet B only, 
and \textit{Chandra}
shows that the B component displays a harder distribution of X-ray
flares. Such flares may repeatedly produce and trap electrons in the extended
magnetosphere at a rate sufficient to sustain the quasi-steady radio 
emission. Given the similar masses
and spectral types of the binary components, it is yet unclear why
their coronal behavior should differ.
Stellar rotation may play an important role, since coronal activity is
roughly proportional to the rotation rate.
Infrared spectroscopy has
recently shown that UV Cet A rotates relatively fast ($v \sin i
= 58$~\kms; \citealt{hinkle03}), but no measurement is yet
available, to our knowledge, for UV Cet B. 
The physical causes responsible for the disparate coronal behaviors of these
two stars remain yet unexplained, however the answer to this
question may provide important information on the generation and sustainment of
magnetic fields in late-type stars.

\acknowledgments
MA and MG acknowledge support from the Swiss National Science Foundation (fellowship 81EZ-67388 and
grant 2000-058827, respectively). SS was supported by SAO grant GO1-2011X. We
are thankful to V.~Kashyap for useful technical information about the LETGS.

\clearpage

\notetoeditor{Please put Figure 2 as a two-column figure, whereas the other
figures should be in one-column size.}

\end{document}